\documentclass[twocolumn,reprint,
 superscriptaddress, amsmath,amssymb, aps, prl]{revtex4-1}
\usepackage{graphicx}
\usepackage{amsmath}
\usepackage{color}
\usepackage{dcolumn}
\usepackage{bm}
\usepackage[
   bookmarks=true,
   colorlinks=true,
   hyperindex=true,
   citecolor=blue,
   linkcolor=blue,
   urlcolor=blue
]{hyperref}

\begin{document}
\title{Defect-free arbitrary-geometry assembly of mixed-species atom arrays}
\author{Cheng Sheng}
\affiliation{State Key Laboratory of Magnetic Resonance and Atomic and Molecular Physics, Wuhan Institute of Physics and Mathematics, Innovation Academy for Precision Measurement Science and Technology, Chinese Academy of Sciences, Wuhan 430071, China}
\author{Jiayi Hou}
\affiliation{State Key Laboratory of Magnetic Resonance and Atomic and Molecular Physics, Wuhan Institute of Physics and Mathematics, Innovation Academy for Precision Measurement Science and Technology, Chinese Academy of Sciences, Wuhan 430071, China}
\author{Xiaodong He}
\email{hexd@wipm.ac.cn}
\affiliation{State Key Laboratory of Magnetic Resonance and Atomic and Molecular Physics, Wuhan Institute of Physics and Mathematics, Innovation Academy for Precision Measurement Science and Technology, Chinese Academy of Sciences, Wuhan 430071, China}
\affiliation{School of Physical Sciences, University of Chinese Academy of Sciences, Beijing 100049, China}
\author{Kunpeng Wang}
\affiliation{State Key Laboratory of Magnetic Resonance and Atomic and Molecular Physics, Wuhan Institute of Physics and Mathematics, Innovation Academy for Precision Measurement Science and Technology, Chinese Academy of Sciences, Wuhan 430071, China}
\author{Ruijun Guo}
\affiliation{School of Information Engineering and Henan Key Laboratory of Laser and Opto-Electric Information Technology, Zhengzhou University, Zhengzhou 450001, China}
\author{Jun Zhuang}
\affiliation{State Key Laboratory of Magnetic Resonance and Atomic and Molecular Physics, Wuhan Institute of Physics and Mathematics, Innovation Academy for Precision Measurement Science and Technology, Chinese Academy of Sciences, Wuhan 430071, China}
\affiliation{School of Physical Sciences, University of Chinese Academy of Sciences, Beijing 100049, China}
\author{Bahtiyar Mamat}
\affiliation{State Key Laboratory of Magnetic Resonance and Atomic and Molecular Physics, Wuhan Institute of Physics and Mathematics, Innovation Academy for Precision Measurement Science and Technology, Chinese Academy of Sciences, Wuhan 430071, China}
\affiliation{School of Physical Sciences, University of Chinese Academy of Sciences, Beijing 100049, China}
\author{Peng Xu}
\email{etherxp@wipm.ac.cn}
\affiliation{State Key Laboratory of Magnetic Resonance and Atomic and Molecular Physics, Wuhan Institute of Physics and Mathematics, Innovation Academy for Precision Measurement Science and Technology, Chinese Academy of Sciences, Wuhan 430071, China}
\author{Min Liu}
\affiliation{State Key Laboratory of Magnetic Resonance and Atomic and Molecular Physics, Wuhan Institute of Physics and Mathematics, Innovation Academy for Precision Measurement Science and Technology, Chinese Academy of Sciences, Wuhan 430071, China}
\author{Jin Wang}
\affiliation{State Key Laboratory of Magnetic Resonance and Atomic and Molecular Physics, Wuhan Institute of Physics and Mathematics, Innovation Academy for Precision Measurement Science and Technology, Chinese Academy of Sciences, Wuhan 430071, China}
\author{Mingsheng Zhan}
\affiliation{State Key Laboratory of Magnetic Resonance and Atomic and Molecular Physics, Wuhan Institute of Physics and Mathematics, Innovation Academy for Precision Measurement Science and Technology, Chinese Academy of Sciences, Wuhan 430071, China}

\date{\today}

\begin{abstract}

Optically trapped mixed-species single atom arrays with arbitrary geometries are an attractive and promising platform for various applications, because tunable quantum systems with multiple components provide extra degrees of freedom for experimental control. Here, we report the first demonstration of two-dimensional $6\times4$ dual-species atom assembly with a filling fraction of 0.88 (0.89) for $^{85}$Rb ($^{87}$Rb) atoms. This mixed-species atomic synthetic is achieved via rearranging initially randomly distributed atoms using a sorting algorithm (heuristic heteronuclear algorithm) which is proposed for bottom-up atom assembly with both user-defined geometries and two-species atom number ratios. Our fully tunable hybrid-atom system of scalable advantages is a good starting point for high-fidelity quantum logic, many-body quantum simulation and forming defect-free single molecule arrays.

\end{abstract}

\maketitle



\maketitle


Single neutral atoms trapped in optical tweezer arrays have been widely used to simulate quantum many-body dynamics~\cite{Bernien2017,Lienhard2018,Keesling2019,leuc2019,Song2021,Scholl2020,Ebadi2020,Kim2020,Bluvstein2021}, form cold molecules~\cite{Liu2018,Liu2019,Zhang2020,He2020}, build quantum computers~\cite{Xia2015,Yang2016,Wang2017,Sheng2018,Guo2020,Levine2019,Graham2019,Madjarov2020,Saffman2016,Zeng2017} and precisely measure optical frequency~\cite{Norcia2019,Young2020} in recent years. Generally, new appealing applications are emerged from more degrees of freedom in this quantum system are experimentally controlled. For instance, as the number of atoms increased to hundreds, quantum simulators based on Rydberg atom arrays can solve quantum many-body problems which are intractable on classical machines~\cite{Scholl2020,Ebadi2020,Bluvstein2021}. Additionally, three-dimensional quantum annealers with the adjustable connectivity of atoms bring prospects to quantum optimization problems~\cite{Kim2020,Song2021}. These various applications are demonstrated on single-species defect-free atom arrays, while mixed-species atom assembly should be a more versatile platform in the future.

For quantum logic, ion system greatly benefit from the use of dual-species ions (one species for data ions and the other species for ancilla ions)~\cite{Schmidt2005,Tan2015,Ballance2015,Negnevitsky2018,Sosnova2021}. Because a decoherence of data ions occurs when photons re-emitted by measurement or the sympathetic cooling process of ancilla ions are absorbed by data ions if they are identical~\cite{Bruzewicz2019}. Akin to trapped ions, the implements of high-fidelity quantum logic in the neutral atom platform will also be impeded by these problems. Therefore, the dual-species neutral atom array is essential to realize some schemes proposed for nondestructive cooling via Rydberg interactions~\cite{Belyansky2019}, nondemolition state measurement with low cross-talk~\cite{Beterov2015} and the surface code for quantum error correction~\cite{Fowler2012,Auger2017}. In addition, substantially different resonant frequencies of the dual-species atom allow individual addressing in atom array with low-cross talk (induced by the focused addressing laser beam when quantum bits spaced closed).

On the other hand, cold molecules are formed via Feshbach resonance of two dual-species atoms in optical lattice with filling fractions of 25$\%$ to 30$\%$ limited by the preparing efficiency of dual-species atom pair in one site (conversion efficiency from atoms to Feshbach molecule can reach about 90$\%$)~\cite{Reichs2017,Moses2015}. In recent years, merging two single atoms in optical tweezers has been developed to be an alternative approach to building single molecules~\cite{Liu2018,Liu2019,Zhang2020,He2020}. Along this way, bottom-up defect-free dual-species atom array could be an excellent starting point to generate ultracold single molecule arrays with high filling fractions.

Multi-species might also provide the flexibility to control the interaction between the atoms for quantum simulations~\cite{Weimer2010,Browaeys2020}. With such versatile applications, dual-species atom array would become a promising platform for quantum computation, quantum simulation and single molecule arrays. So far, single-species defect-free atom arrays have been well prepared from one-dimension to three-dimension~\cite{Kim2016,Endres2016,Barredo2016,Lee2017,Barredo2018,Kumar2018,Brown2019,Mello2019,Schymik2020,Sheng2021}. However, the generation of mixed-species defect-free atom arrays have not been reported yet for both bottom-up and top-down approaches.

\begin{figure*}[htbp]
\centering
\includegraphics[width=18cm]{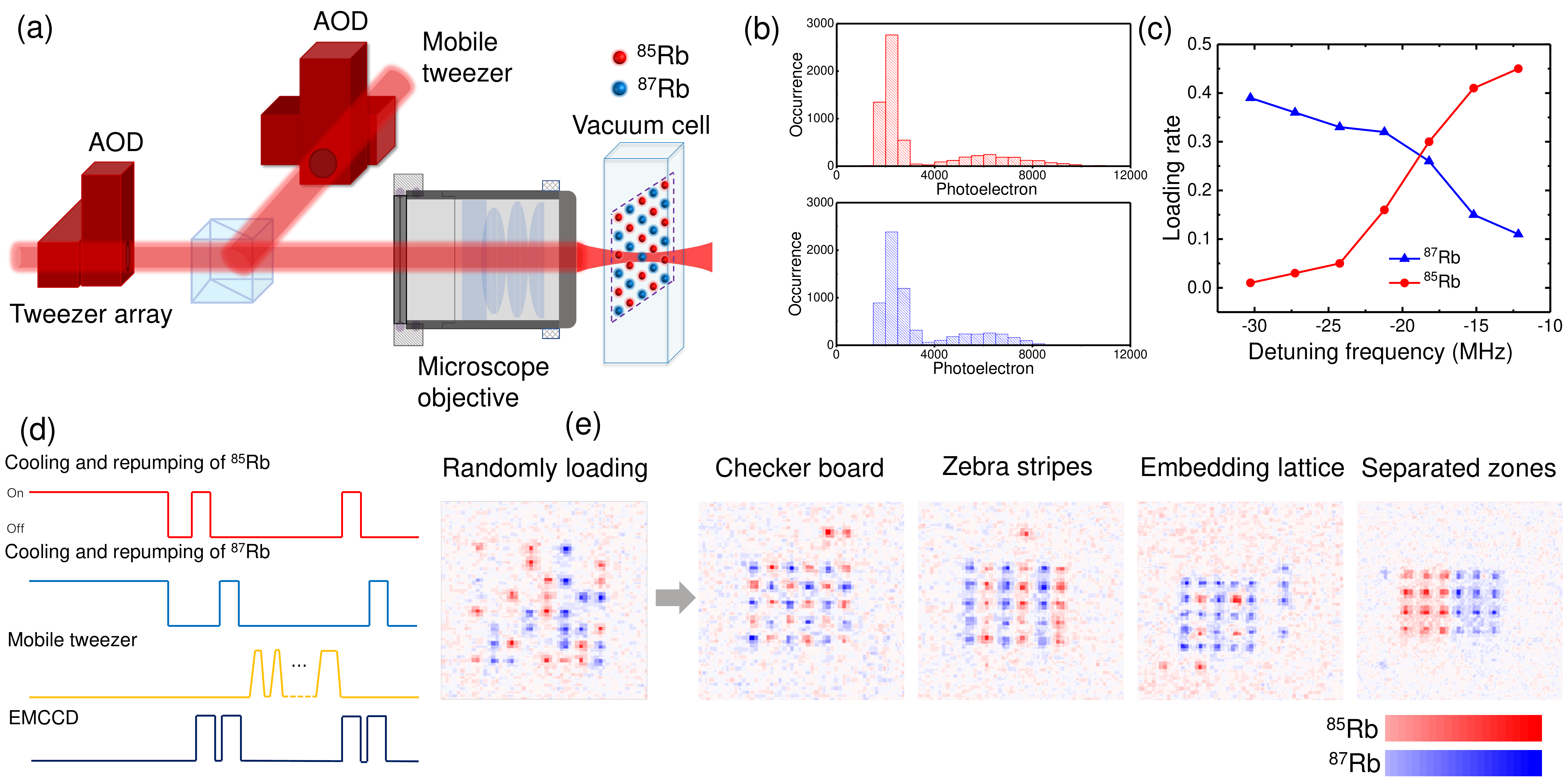}
\caption{(color online). Protocol for creating mixed-species atom array. (a) Simplified schematics of the experimental setup. (b) Fluorescence count distribution for 64 tweezers during 80 ms of exposure of $^{85}$Rb (top) and $^{87}$Rb (down) atoms. (c) The loading rates of $^{85}$Rb and $^{87}$Rb as a function of the frequency of $^{85}$Rb cooling beam. The frequency of $^{87}$Rb cooling beam is fixed to 15 MHz detuned by $|5s_{1/2}, f=2\rangle\rightarrow|5p_{3/2}, f=3\rangle$ and the loading ratios are tuned by the frequency of $^{85}$Rb cooling beam detuned by $|5s_{1/2}, f=3\rangle\rightarrow|5p_{3/2}, f=4\rangle$. The optical intensity of $^{85}$Rb and $^{87}$Rb cooling (repumping) beams are almost the same. The tunable loading ratio is essential for the atom arrays with imbalanced two-species atom numbers such as ``embedding lattice" configuration. (d) Pulse diagram of the experimental sequence. (e) Images of various defect-free mixed-species atom arrays. Each one is synthetized by two images of $^{85}$Rb and $^{87}$Rb. The distance of the neighboring sites is 5 $\mu$m.}
\label{fig:fig1}
\end{figure*}

Here, we present the realization of preparing two-dimensional heteronuclear atom ($^{85}$Rb and $^{87}$Rb) arrays with arbitrary geometries. Our first step is generating a dual-species initial atom array with controllable ratios of two-species atom number. Next, the defect-free atom assembly is formed by rearranging the randomly distributed atoms using a heuristic heteronuclear algoritnm (HHA) which is proposed for defect-free dual-species atom arrays with user-defined geometries. The measured filling fractions of $6\times4$ atom assembly are 0.89 (0.88) for $^{87}$Rb ($^{85}$Rb) atoms mainly limited by the atom survival probability in the optical tweezers. After the sorting atom process, the configuration entropy of the dual-species atom system is lowered by a factor of 5.

Our experimental setup is shown schematically in Fig.~\ref{fig:fig1}(a) (see details in Ref.~\cite{Sheng2021}). In brief, the 808-nm $8\times8$ tweezer array (with trap depth of $U_{0}/k_{B}\approx$0.8 mK for each tweezer, where $k_{B}$ is the Boltzmann constant) is generated by an acousto-optic deflector (AOD) driven by a multi-tone radio-frequency (rf) and the 830-nm mobile tweezer (2.3 mK) is created by another AOD driven by a dynamically controlled rf (both 808-nm and 830-nm tweezers are red-detuned optical dipole traps for $^{85}$Rb and $^{87}$Rb). The dual-species single atoms are stochastically loaded into an $8\times8$ tweezer array from a magneto-optical trap (the magneto-optical traps of $^{85}$Rb and $^{87}$Rb are overlapped) simultaneously with a total loading rate about 50$\%$. Unlike the single-species atom loading, light-assisted collisions between a heteronuclear atoms occur during the loading process in 950 ms. The two-species atom loading ratio is controlled by the frequency of $^{85}$Rb cooling laser with the fixed cooling parameters of $^{87}$Rb as shown in Fig.~\ref{fig:fig1}(c). The images of $^{85}$Rb and $^{87}$Rb atom fluorescence are taken by an electron multiplying charge-coupled device camera to reveal their initial positions in the atom arrays. The photoelectron count thresholds as shown in Fig.~\ref{fig:fig1}(b) and the regions of interest of $^{85}$Rb and $^{87}$Rb for every tweezer are respectively calibrated via automatically analyzing a series of atom array images. Next, according to the two initial images of $^{85}$Rb and $^{87}$Rb detected respectively by their fluorescence and the user-defined target sites, the HHA is used to calculate the sorting-atom path of the mobile tweezer. The sorting-atom path is composed of a list of voltage sequences to control the optical intensity and X and Y positions of the mobile tweezer (a rearrangement for one atom consists of extracting the atom by the mobile tweezer with a deeper trap depth, moving the mobile tweezer along a calculated path, and releasing the atom onto a target site). Finally, after the rearrangement for all atoms accomplished, the new images of $^{85}$Rb and $^{87}$Rb atom array are acquired as illustrated in Fig.~\ref{fig:fig1}(e). The total experimental process sequence is depicted in Fig.~\ref{fig:fig1}(d).

\begin{figure*}[htbp]
\centering
\includegraphics[width=18cm]{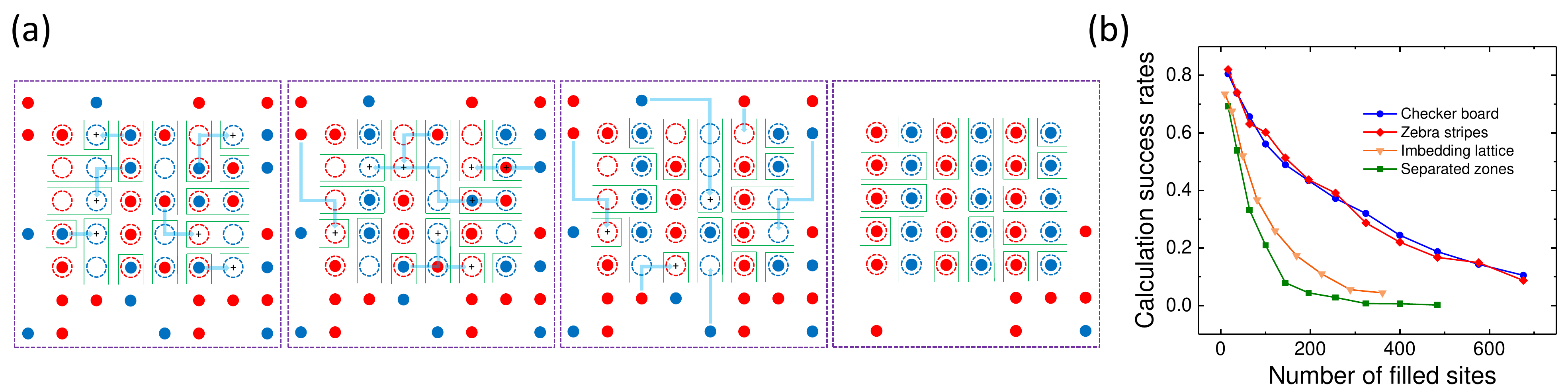}
\caption{(color online). Assembling of a mixed-species atom array using the HHA. (a) Schematic diagrams of the HHA. The sorting process is from left to right. The blue (red) disks denote single atoms of $^{87}$Rb ($^{85}$Rb) and blue (red) dash line circles denote the target sites of $^{87}$Rb ($^{85}$Rb). The green regions represent the routes connected to the reservoir tweezers and the sites with ``+" symbols are the innermost sites in the routes to be filled. (b) Calculation success rates of the HHA as the number of filled sites increased. Each point is the average value of the simulation by rearranging a randomly loading atom array for 1000 times.}
\label{fig:fig2}
\end{figure*}


A sorting-atom algorithm well-suited for dual-species atom arrays is much more complicated than single-species atom algorithm. To solve this problem, we propose a sorting-atom algorithm (HHA) for bottom-up atom assemblers with user-defined geometries and atom number ratios. Here, we give a brief introduction of the working principle as shown in Fig.~\ref{fig:fig2}(a). The HHA is heuristic algorithm aiming at efficiently sorting dual-species atoms (both of them can be trapped in every optical tweezer of the array and atoms are moved along the site links) into target configurations with good scalability. In the first step, target sites to be filled are classified into three types by the HHA. Type-1 sites are the finished sites which have already been filled with target atoms after the initially loading process. Atoms in type-1 sites will not be moved during the whole rearrangement. The remained target sites are unfinished sites. Type-2 sites are filled with misplaced atoms (for example, a $^{85}$Rb target site is occupied with a $^{87}$Rb atom) and type-3 sites are the empty sites. Next, the unfinished sites are divided into several groups as the green regions show in Fig.~\ref{fig:fig2}(a). The groups are defined as routes to connect the unfinished sites to the reservior atoms in non-target sites (also called the outside-region). A route may be composed of a number of branches (which are not illustrated in Fig.~\ref{fig:fig2}(a)) if the array size is large and the whole routes should cover all unfinished sites. There also two types of source atoms used to fill the unfinished sites. The type-M source atoms are the misplaced atoms in type-2 sites and the type-R source atoms are the reservior atoms in the outside-region.

After the classification and grouping, the unfinished sites are to be filled in order. The innermost type-3 sites (the site with the longest distance from the outside region) in each route are filled with nearest type-M source atom firstly (the source atoms should be directed moved to target sites without any obstacle atoms blocking the path). When this process are accomplished, some innermost type-3 sites are converted to the type-1 sites, meanwhile some type-2 sites are turned into type-3 sites (because the misplaced atoms in type-2 sites are move to type-3 sites). Due to the occurrence of new type-3 sites, a repeated process is carried out to fill the new type-3 sites. After several cycles, most of the innermost type-2 sites become type-1 sites. Next, the second (third and so on) innermost sites are filled in the same way until the type-M source atoms are almost used out (a few of type-M source atoms can not be used are moved to the outside-region). There are still remained some type-3 sites due to a lack of type-M source atoms. These sites then turn to search the nearest type-R source atoms to become unit filling. Finally, the rest of misplaced atoms in type-2 sites are transported to the empty reservior tweezers and the remained type-3 sites (caused by removing misplaced atoms) are filled with type-R source atoms. The flowchart of the HHA is shown in~\cite{S1}.

Most of the time, we can acquire the sorting path of preparing a dual-species atom array using the HHA as depicted in Fig.~\ref{fig:fig2}(a). However, sometimes the HHA can not provide a sorting-atom scheme if the initial atom array in some special configuration. One case is that during the process of transporting the type-R source atoms to the target sites the obstacle atoms (blocked by other atoms) can not be directly moved to reservior tweezers. The other case is that the unfinished sites are surrounded by the finished sites, but the possibility of this case is low when the array size is not large (an occurrence probability of 1.25$\%$ for a $5\times6$ atom array in ``zebra stripes" configuration). In our experiment, we reload the atom array if the HHA can not calculate the sorting path. The simulation success rates as a function of filled sites are as shown in Fig.~\ref{fig:fig2}(b). The decay of calculation success rate can be mitigated by using more sophisticated sorting-atom approaches to deal with the above case in future. Moreover, we tried to improve the rates by modifying the filling order of the HHA, we find that the modified HHA can increase the rates of ``separated zones" configuration but decrease the rates of other types. Therefore, choosing a algorithm appropriate for the target array configuration is necessary. Additionally, we should note that the filling fraction is affected by the number of moves~\cite{Schymik2020,Sheng2021} rather than the calculation success rate. On account of this, HAA is designed to provide the number of moves as few as possible~\cite{S2} to make the HHA suited for efficiently preparing dual-species atom arrays on larger scale.


\begin{figure}[htbp]
\centering
\includegraphics[width=8.6cm]{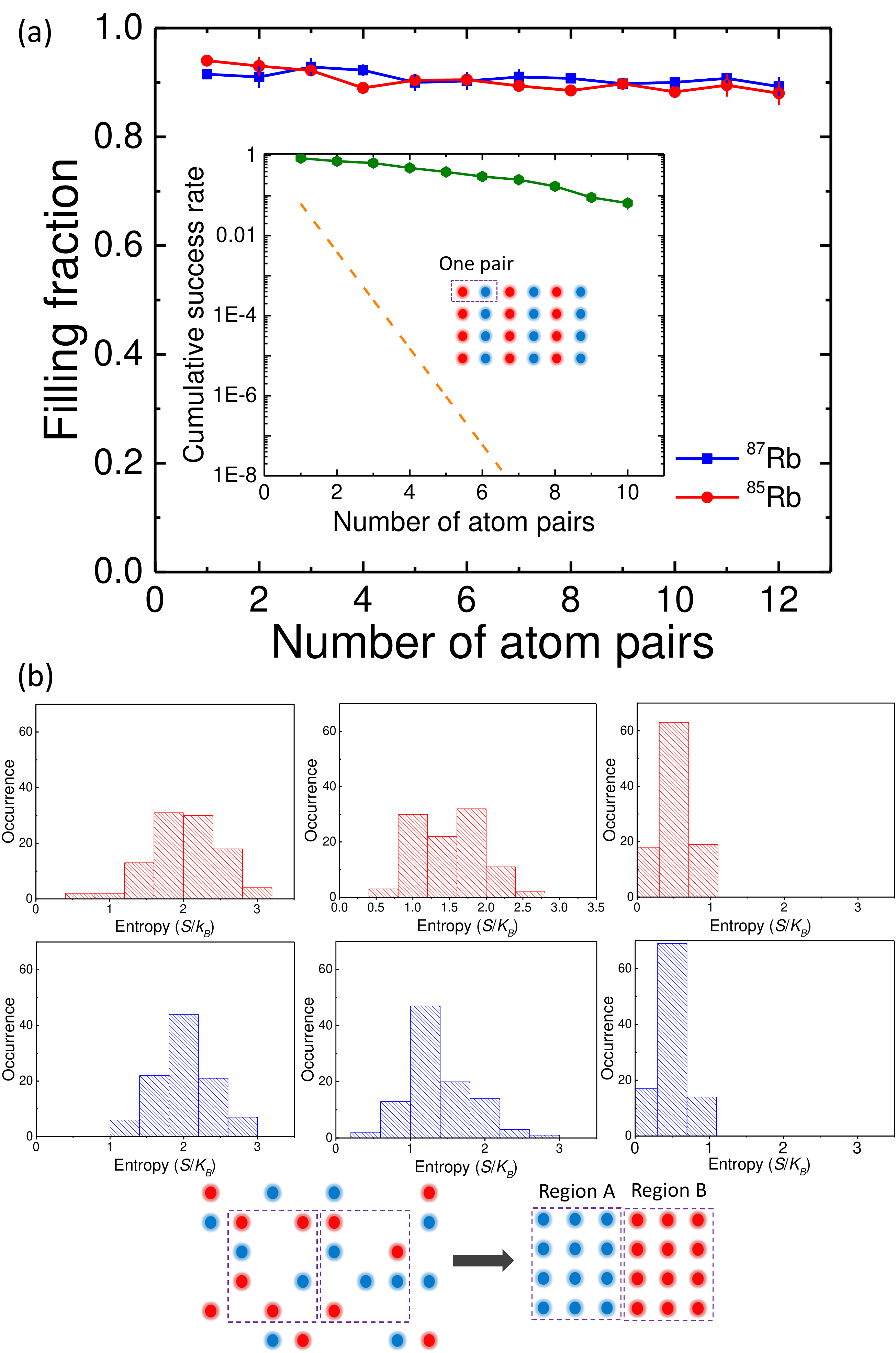}
\caption{(color online). Filling fraction and configuration entropy of the dual-species atom array. (a) Filling fraction of $^{85}$Rb and $^{87}$Rb as a function of the number of atom pairs. Inset: Green circles show the cumulative success rate of obtaining a defect-free atom array with several atom pairs; the dashed line shows the result corresponding to random loading for dual-species atoms. (b) Configuration entropy decreasing in the process of rearrangement. Each entropy distribution is measured by repeating the experiment for 100 times. The left (right) one is the result of initial (final) atom array and the up (down) one is the result of region A (B). The middle one represents the entropy with half of the total transport moves. The values of entropy during the sorting process in region A and B are 2.0(5), 1.5(4), 0.4(3)$k_{B}$ and 2.0(4), 1.4(4), 0.4(2)$k_{B}$ respectively.}
\label{fig:fig3}
\end{figure}


Analyzing 100 repetitions of the experiment for mixed-species atom arrays in ``zebra stripes" configuration, the filling fractions of $^{85}$Rb and $^{87}$Rb target sites are measured as the number of atom pairs is increased (Fig.~\ref{fig:fig3}(a)). We achieve a preparation of 12 dual-species atom pairs with a filling fraction of 0.89(2) for $^{87}$Rb and 0.88(2) for $^{85}$Rb after one arrangement cycle. As the inset shows in Fig.~\ref{fig:fig3}(a), our method significantly improved the cumulative success rates of the defect-free mixed-species atom array ($P_{n}=2^{-2n}$ for dual-species array and $P_{n}=2^{-n}$ for single-species array with randomly loading, if the loading ratios are 25$\%$ for both two-species atoms and $n$ is the number of filled sites). These filling fractions are mainly limited by the atom lifetime in the tweezer array (92$\%$ $^{87}$Rb and 94$\%$ $^{85}$Rb atoms are survived when the rearrangement process is not implemented). The atom loss is mainly induced by the heating in the probing atom process and the background gas collisions. Another factor that affects the filling fraction is the atom transport efficiency (typically about 98$\%$ in our experiment), but the experimental imperfect in transport process can be mitigated by rearranging the atom array for several cycles. The array size can be further increased to several hundreds only requiring sufficient laser power for reliable atom trapping with no more changes in the other hardware.

Additionally, we measure the configuration entropy during rearrangement in a frame of Maxwell's demon like as depicted in Fig.~\ref{fig:fig3}(b). The entropy per particle in region A(B) is defined as~\cite{S3}
\begin{equation}\label{eq1}
S=-\frac{k_{B}}{n+m}[n\ln{n}+m\ln{m}+(1-n-m)\ln{(1-n-m)}],
\end{equation}
where n(m) is the filling fraction of $^{85}$Rb($^{87}$Rb) atoms. In this microscopic system with the initial configurational disorder, the entropy is reduced by the sorting atom process. A factor of 5 for lowering the entropy is achieved. In fact, an information entropy is increased when our demon acquires the atom occupancy of the whole system and map out a plan to act on all the particles. Therefore, the second law of thermodynamics is not violated by the lowering entropy process. The configuration entropy of the final atom array is not zero because each region can not reach unit filling due to the experimental imperfections. Additionally, the heating entropy can also be decreased if we adopt the method of dual-species Raman sideband cooling in our previous experiment~\cite{Wang2019}.

Dual-species atom arrays in specific patterns as illustrated in Fig.~\ref{fig:fig1}(e) have their designed purpose. For instance, ``checker board" configuration is suited for surface code (one species for data qubits, the other one for measurement qubits) or scheme proposed for quantum nondemolition state measurements based on Rydberg interaction. ``zebra stripes" configuration is prepared for building single molecule arrays. The neighboring two columns of two-species atoms can be merged into one columns of tweezers with each one trapping two atoms. The merging process could be achieved by dynamically tuning the tones of rf (similar to the approach in Ref.~\cite{Endres2016}). The atom pairs then can be converted to the single molecules by the means of photoassociation, Feshbach or spin-motion coupling.

Atom-by-atom assembly of defect-free dual-species atom array forms a versatile scalable platform not only for multiqubits with quantum error correcting or the atom-molecule system with full control, but also has perspective to build an atom system mixed of bosons and fermions with determined particle number and complex spin patterns using a dual-element atom array~\cite{S4} in the future. For the experimental realization, preparing dual-element atom pairs in three-dimensional motional ground state~\cite{Liu2019} and simultaneous cooling all atoms in the array at once via Raman sideband cooling have been demonstrated~\cite{Kumar2018,Lorenz2021}. Coherent atom tunneling effect has also been observed between two optical tweezers~\cite{Kaufman2014}.

In summary, we present a novel platform of defect-free dual-species atom assembly with arbitrary geometries and good scalability. This approach opens promising paths to realizing high-fidelity quantum logic, ultracold single molecule arrays and quantum simulation with complex degrees of freedom.

\section{acknowledgments}
This work was supported by the National Key Research and Development Program of China under Grant Nos.2017YFA0304501, 2016YFA0302800, and 2016YFA0302002, the National Natural Science Foundation of China under Grant No.11774389, No.12004395, the Strategic Priority Research Program of the Chinese Academy of Sciences under Grant No.XDB21010100 and the Youth Innovation Promotion Association of CAS under Grant No.2019325, No.2017378.


\clearpage
\onecolumngrid
\noindent
\begin{center}
\large{\bf Supplemental Material: Defect-free assembly of mixed species atom arrays with arbitrary geometries \\}
\end{center}
\date{\today}

\section{The flowchart of the HHA}
\begin{figure*}[htbp]
\centering
\includegraphics[width=18cm]{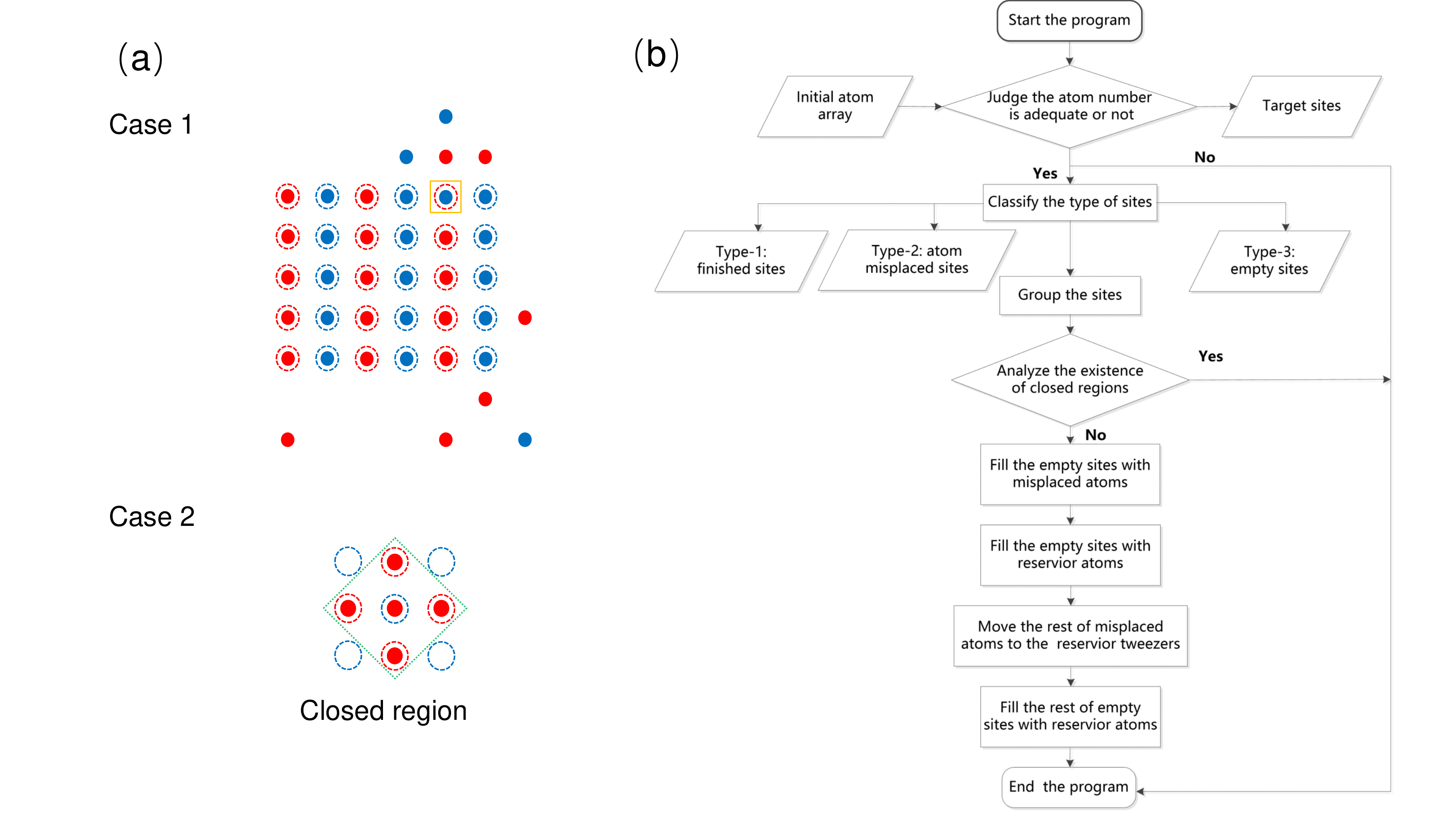}
\caption{(color online). (a) Two cases can not be solved by HHA. Case-1: the obstacle atoms that blocks the sorting path is also blocked by other atoms. Case-2: The unfinished sites are surrounded by the finished sites. (b) The flowchart of the HHA.}
\label{fig:fig1s}
\end{figure*}
To further improve the calculation success rate, more complicated sorting-path should be designed to deal with the above cases.

\section{The number of moves of the HHA}
\begin{figure*}[htbp]
\centering
\includegraphics[width=18cm]{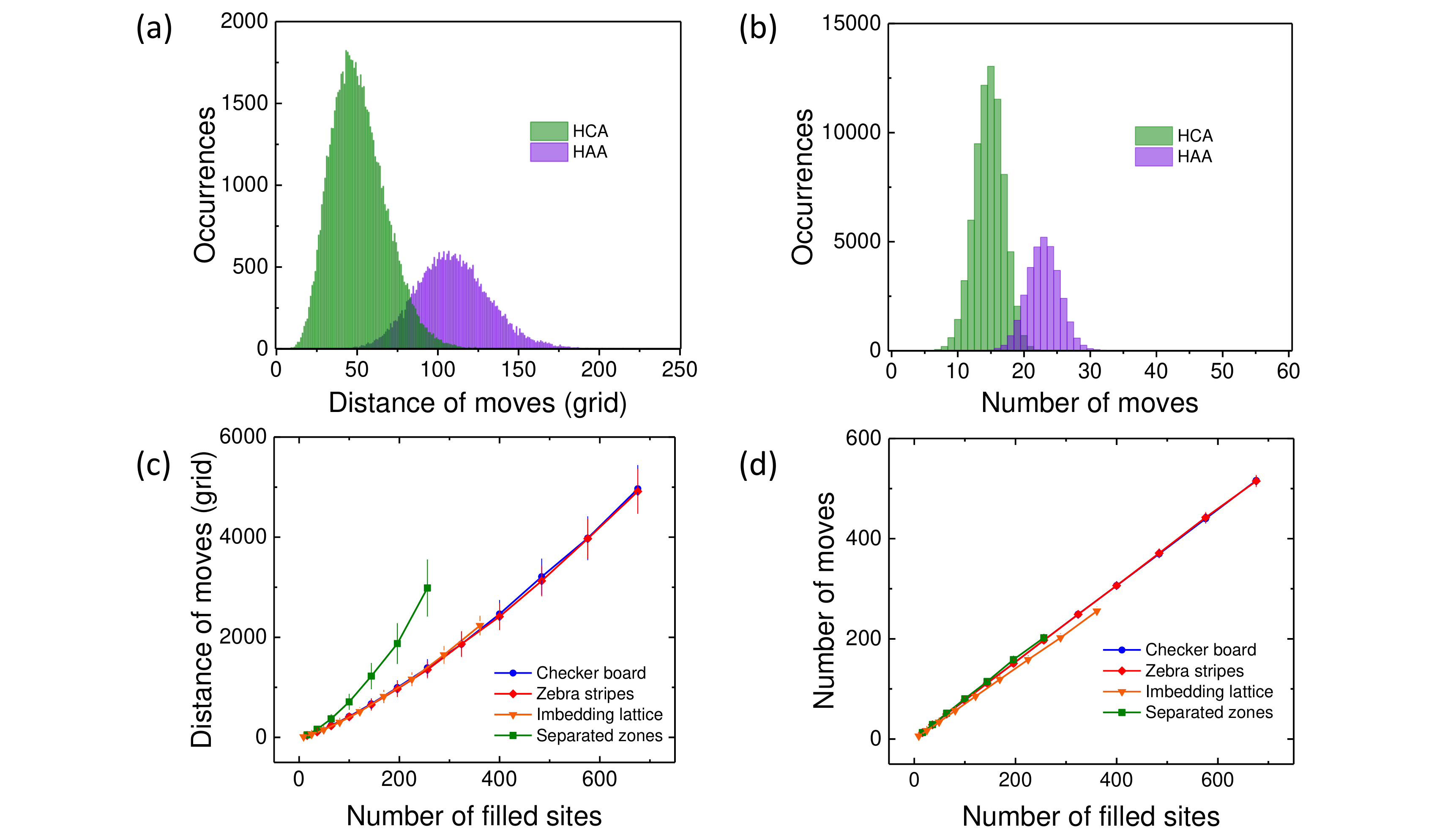}
\caption{(color online). (a) The distance of moves of the HHA compared with HCA. One grid represents the distance between two adjacent sites. The simulation results are repeated by 100000 times for a $5\times6$ single-species atom array and $5\times6$ dual-species atom array in ``zebra stripes" configuration. The smaller total area of the distribution of the HHA is caused by the calculation success rates. (b) The number of moves of the HHA compared with the HCA. (c) The distance of moves as a function of the number of filled sites for dual-species atom array. (d) The number of moves as a function of the number of filled sites for dual-species atom array.}
\label{fig:fig2s}
\end{figure*}
The filling fraction of a rearranged atom array is closely related to the number of sorting-atom moves. The less number of moves is provided by an algorithm, the higher filling fraction of the final atom assembly is. Therefore, the HHA has also been designed to reduce the number of moves. Compared with heuristic cluster algorithm (HCA), Fig.~\ref{fig:fig2s}(a) and Fig.~\ref{fig:fig2s}(b) show that the HHA costs more distance and number of moves than HCA (HCA is algorithm with near fewest sorting-atom moves for single-species atom array which has been described in Ref.~\cite{Sheng2021}). This is mainly resulted from the transporting of misplaced atoms which is not occurred in the single-species atom array. Additionally, after the randomly loading process, only about 25$\%$ target sites of dual-species atom array are initially finished, while 50$\%$ sites are fully filled for single-species atom array as usual. Nevertheless, the number of moves of the HHA is still optimized, even smaller than some algorithms of single-species atom array.

The HHA gives a number of moves linear in the number of the filled sites in three configuration as shown in Fig.~\ref{fig:fig2s}(c). This indicates that the HHA is well-suited for large-scale dual-species atom array~\cite{Schymik2020,Sheng2021}. The distances of moves of the HHA are differed from the atom array configuration as illustrated in Fig.~\ref{fig:fig2s}(d).

\section{Schemes for dual-element atom arrays}

\begin{figure*}[htbp]
\centering
\includegraphics[width=15cm]{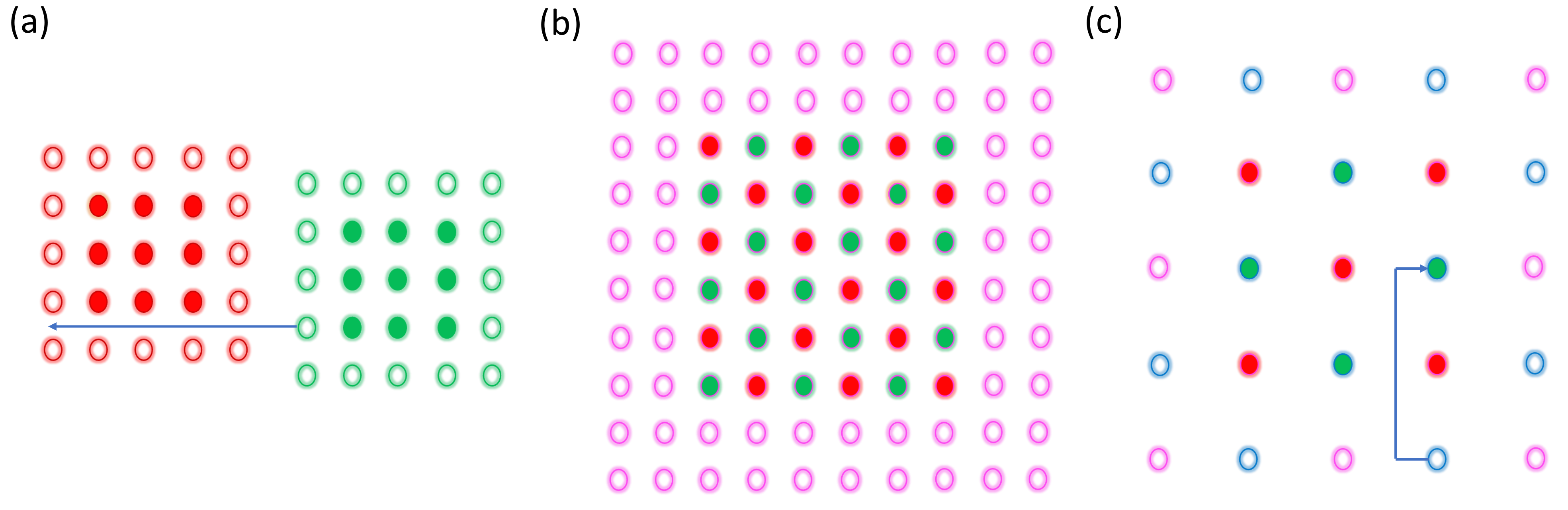}
\caption{(color online). Schemes for generating dual-element atom arrays. (a) Dual-element atom array in ``zebra stripes" configuration. The green and red disks (circles) denote dual-element atoms (tweezers). (b) Dual-element atom array that can be prepared by the HHA. (c) Atoms transported along the path between adjacent atom rows.}
\label{fig:fig3s}
\end{figure*}

Generally, single neural atoms trapped in optical tweezers are alkali metal atoms and alkaline-earth metal atoms. That the situation we discuss here is atoms trapped in red-detuned optical dipole trap. For the simplest case, to prepare a single molecule array formed by dual-element atoms, we should firstly prepare two single-species atom array spatially separated as shown in Fig.~\ref{fig:fig3s}(a). These two atom arrays then move into a dual-element atom array in ``zebra stripes" configuration. Finally, dual-element atoms in two neighboring columns can be merged into one column to form a single molecule array.

To generate a dual-element atom array in arbitrary configuration (such as the ``checker board" configuration as shown in Fig.~\ref{fig:fig3s}(b)), there are two cases should be discussed respectively. For the first case, the optical tweezer  a red-detuned optical dipole trap for both two-species atom. We choose a tweezer array with an appropriate wavelength (for example, 960-nm tweezer is red-detuned optical dipole trap for both Rb and Cs atoms) or overlap two tweezer arrays with an appropriate intensity ratio for reliably trapping both two-element atoms in each tweezer. Then we can adopt the same method of forming atom assembly of $^{87}$Rb and $^{85}$Rb atoms to prepare a dual-element atom array. The second case is that the red-detuned optical tweezer for one of the two-species atom is a blue-detuned optical tweezer for the other species atom. A feasible approach to achieving dual-element atom array is moving the atoms across the gaps of two sites as shown in Fig.~\ref{fig:fig3s}(c) and the sorting-atom algorithm is much simpler than the HHA. However, the atom array configurations are constrained with the tweezer arrays and less atoms can be trapped in such a sparse structure.

\section{Configuration entropy of the multiple-species atom arrays}
The definition of the configuration entropy of the atom arrays is derived from the Shannon entropy. For the case of single-species atom arrays, the void site and the filled sites are the two components of the atom array. The configuration entropy per atom is given by~\cite{Reichs2017,Kumar2018}
\begin{equation}\label{eq2}
S=-\frac{k_{B}}{n}[n\ln{n}+(1-n)\ln{(1-n)}],
\end{equation}
where n is the average filling fraction of the atom arrays. It is obvious that the entropy is only concerned with the ratio of the void (or the filled) sites to the total sites rather than the geometries of the atom arrays. The configuration entropy per atom can be also applied to the case of multi-species atom arrays with the definition as
\begin{equation}\label{eq3}
S=-\frac{k_{B}}{\sum_{i=1}^Nn_i}[\sum_{i=1}^Nn_i\ln{n_i}+(1-\sum_{i=1}^Nn_i)\ln{(1-\sum_{i=1}^Nn_i)}],
\end{equation}
where N is the number of the atom species and $n_i$ represents the filling fraction of the species of $i$ atoms.
\end{document}